# Constrained Role Mining

Carlo Blundo
DI, Università degli studi di Salerno
84084 Fisciano (SA), Italy
cblundo@unisa.it

Stelvio Cimato
DTI, Università degli studi di Milano
26013 Crema (CR), Italy
stelvio.cimato@unimi.it

## ABSTRACT
Role Based Access Control (RBAC) is a very popular access control model, for long time investigated and widely deployed in the security architecture of different enterprises. To implement RBAC, roles have to be firstly identified within the considered organization. Usually the process of (automatically) defining the roles in a bottom up way, starting from the permissions assigned to each user, is called *role mining*. In literature, the role mining problem has been formally analyzed and several techniques have been proposed in order to obtain a set of valid roles.

Recently, the problem of defining different kind of constraints on the number and the size of the roles included in the resulting role set has been addressed. In this paper we provide a formal definition of the role mining problem under the cardinality constraint, i.e. restricting the maximum number of permissions that can be included in a role. We discuss formally the computational complexity of the problem and propose a novel heuristic. Furthermore we present experimental results obtained after the application of the proposed heuristic on both real and synthetic datasets, and compare the resulting performance to previous proposals.

## 1. INTRODUCTION
Complex organizations need to establish access control policies in order to manage access to restricted resources. A simple way to accomplish this task is to collect set of permissions in roles and then assign roles according to the responsibilities and qualifications of each employee. The Role Based Access Control (RBAC) is a well known paradigm to define and organize roles and permissions in an efficient way. Introduced in the early '90 years [4, 7],such a paradigm has been investigated for long time and has become recently used in different commercial systems to manage identities and accounts [2]. The goal of RBAC is to collect set of permissions in roles and define a complete and efficient set of roles that can be assigned to users in order to access restricted resources. The advantage is that access control can be centralized and decoupled from users and the costs and the overhead of the security management can be reduced.

The correct definition of the set of roles which satisfies the needs of the organization is the most difficult and costly task to be performed during the implementation of a RBAC system. Such an activity is often referred to as *role engineering* and includes the correct identification of roles from the current structural organization of the enterprise. Mainly this task, i.e. the extraction of a complete and efficient set of roles, can be performed using two approaches: *top-down* or *bottom-up* role engineering. In the first case, roles are defined after that the functionalities of the organization have been well defined and studied, and elicitation activities have been performed. The top down approach usually is labor intensive and involves a large amount of work and time done by humans especially in large enterprises with a large number of business processes, as reported in some case study are available in the literature [18]. On the other hand, the bottom-up process, often denoted also as *role mining* starts from the analysis of the current set of permissions assigned to users, and tries to aggregate them in order to extract and define roles. Obviously, hybrid approaches can exist in which both directions, top-down and bottom-up, can be used in subsequent steps of the analysis in order to refine the returned set of roles.

Bottom-up approach to role mining has been more investigated, since many techniques borrowed from data mining can be applied in an automatic way to the existing configuration of user-permission assignments. A RBAC system can be easily constructed in this way and a starting set of roles can be fastly generated. The problem with such an approach, is that the quality and the number of returned roles often are not so good, since no semantics is taken into consideration when the role mining process is started. In many situations the returned set of roles might not match any functional organization within the analyzed enterprise and the existing business processes might not be adequately represented. An accurate analysis of the returned results is needed to better tune the retrieved representation of roles to the current organizational requirements of the enterprise. A formal definition of the Role Mining Problem (RMP) and some of its variants has been given and deeply analyzed in [23]. There, the NP-completeness of the (decisional) RMP problem has been proved, and the formulation of RMP as a graph covering problem has been done in [6, 26].

The problem of imposing constraints on the set of roles resulting after the role mining process has been considered in the past. Statically or dynamically mutually exclusive roles constraints have been included in RBAC models [19] and in the NASI/NIST standards [7]. According to these constraints, for examples, no user can be assigned contemporary a given set of mutually exclusive roles, or no user can activate simultaneously a set of roles in the same session. Such constraints are often used as mechanisms to achieve separation of duty principles, preventing one user from having a large number of roles to complete a task, or imposing restrictions on the minimum number of users needed to perform a critical activity [14].

Recently, a simple classification of the constraints on the cardinality of the number of roles, users and permissions for a given RBAC system has been proposed [15]. The first heuristic taking into account a cardinality constraint on the number of permissions contained in a role has been proposed by [13]. In its work, however, the proposed results have been compared only on other heuristics which were not able to consider constrained roles. In this work we propose a novel heuristic for mining RBAC roles under cardinality constraint. The algorithm is based on a previous proposal [1], where an initial candidate role set was constructed by considering one role for each user on the basis of the current assignment of permissions. The role set is then refined and updated by eliminating the roles obtained as union of other roles already included in the set and ensuring that the cardinality constraint is respected. Finally, an optimization of the role set is performed by running a lattice reduction procedure, previously described in [6]. The resulting procedure is very efficient in terms of computation time and quality of returned role set. To this aim we present the results obtained by running our heuristics on different datasets, some available over the network, some artificially created. The results are compared with our implementation of the algorithm presented in [13] and analyzed in terms of the metrics presented in [17].

The remainder of this paper is organized as follows. In the next section we discuss related works. Section 2 contains the preliminary concepts needed to define the constrained role mining problem and the discussion on its complexity. In section 4 we introduce our heuristics and compare the solution with related work in section 5. Finally Section 6 presents our conclusions and ongoing work.

## 2. CONSTRAINED RMP

In this section we introduce the CONSTRAINED ROLE MINING PROBLEM and analyze its computational complexity showing that it is NP-complete while its optimization version is NP-hard.

### 2.1 Basic Definitions

The notation we use is borrowed from the NIST standard for *Core Role-Based Access Control* (Core RBAC) [7] and it is adapted to our needs.

We denote with $\mathcal{U} = \{u_1, \ldots, u_n\}$ the set of users, with $\mathcal{P} = \{p_1, \ldots, p_m\}$ the set of permissions, and with $\mathcal{R} = \{r_1, \ldots, r_s\}$ the set of roles, where, for any $r \in \mathcal{R}$, we have $r \subseteq \mathcal{P}$. We also define the following relations. $\mathcal{URA} \subseteq \mathcal{U} \times \mathcal{R}$ is a many-to-many mapping *user-to-role* assignment relation. $\mathcal{RPA} \subseteq \mathcal{R} \times \mathcal{P}$ is a many-to-many mapping *role-to-permission* assignment relation. $\mathcal{UPA} \subseteq \mathcal{U} \times \mathcal{P}$ is a many-to-many mapping *user-to-permission* assignment relation. $\mathcal{RH} \subseteq \mathcal{R} \times \mathcal{R}$ is a partial order over $\mathcal{R}$, which is called a *role hierarchy*. When $(r_1, r_2) \in \mathcal{RH}$, we say that the role $r_1$ is senior to $r_2$.

When needed, we will represent the assignment relations by binary matrices. For instance, by UPA we denote the $\mathcal{UPA}$'s matrix representation. The binary matrix UPA satisfies UPA$[i][j] = 1$ if and only if $(u_i, p_j) \in \mathcal{UPA}$. This means that user $u_i$ possesses permission $p_j$. In a similar way, we define the matrices URA, RPA, and RH.

Given the $n \times m$ users-to-permissions assignment matrix UPA, the *role mining problem* (see [23], [6], and [8]) consists in finding an $n \times k$ binary matrix URA and a $k \times m$ binary matrix RPA such that, UPA $=$ URA $\otimes$ RPA, where, the operator $\otimes$ is such that, for $1 \leq i \leq n$ and $1 \leq j \leq m$, UPA$[i][j] = \bigvee_{h=1}^{k}($URA$[i][h] \wedge $RPA$[h][j])$. Therefore, in solving a role mining problem, we are looking for a factorization of the matrix UPA. The smallest value $k$ for which UPA can be factorized as URA $\otimes$ RPA is referred to as the *binary rank* of UPA. A *candidate* role consists of a set of permissions along with a user-to-role assignment. The union of the candidate roles is referred to as *candidate role-set*. A candidate role-set is called *complete* if the permissions described by any UPA's row can be exaclty *covered* by the union of some candidate roles. In other words, a candidate role-set is complete if and only if it is a *solution* of the *equation* URA $\otimes$ RPA. In this paper we consider only complete candidate role-set.

By adding a contraint $t$ on the number of permissions that can assigned to any roles, the $t$-*constrained role mining problem* can be defined, as follows. Given an $n \times m$ users-to-permissions assignment matrix UPA and a positive integer $t > 1$, find an $n \times k$ binary matrix URA and a $k \times m$ binary matrix RPA such that UPA $=$ URA$\otimes$RPA and, for any $1 \leq i \leq k$, one has $|\{j : $RPA$[i][j] = 1\}| \leq t$. The computational complexity of the above define problem will be discussed in the next section.

### 2.2 NP-Completeness

The computational complexity of the ROLE MINING PROBLEM (and of some of its variants) was considered in several papers (see, for instance, [23], [3], [6], and [24]). In this section we define the decisional version of the $t$-CONSTRAINED ROLE MINING PROBLEM and we show that it is NP-complete (its optimization version is NP-hard). Next we recall the decisional version of the ROLE MINING PROBLEM.

PROBLEM 2.1. (ROLE MINING DECISION PROBLEM) *Given a set of users $\mathcal{U}$, a set of permissions $\mathcal{P}$, a user-permission assignment $\mathcal{UPA}$, and a positive integer $k < \min\{|\mathcal{U}|, |\mathcal{P}|\}$, are there a set of roles $\mathcal{R}$, a user-to-role assignment $\mathcal{URA}$, and a role-to-permission assignment $\mathcal{RPA}$ such that $|\mathcal{R}| \leq k$ and* UPA $=$ URA $\otimes$ RPA*?*

In [23] it was shown that Problem 2.1 is NP-complete. This has been proved by reducing it to the SET BASIS DECISION

PROBLEM (problem SP7 in Garey and Johnson's book [9]) which was shown to be NP-complete by Stockmeyer in [22].

The decisional version of the *t-constrained role mining problem* can be defined, as follows.

PROBLEM 2.2. (*t*-CONSTRAINED ROLE MINING DECISION PROBLEM) *Given a set of users* $\mathcal{U}$, *a set of permissions* $\mathcal{P}$, *a user-permission assignment* UPA, *and two positive integers* $t$ *and* $k$, *with* $t > 1$ *and* $k < \min\{|\mathcal{U}|, |\mathcal{P}|\}$, *are there a set of roles* $\mathcal{R}$, *a user-to-role assignment* $\mathcal{URA}$, *and a role-to-permission assignment* $\mathcal{RPA}$ *such that* $|\mathcal{R}| \leq k$, UPA = URA $\otimes$ RPA, *and, for any* $r \in \mathcal{R}$, $|r| \leq t$?

To prove that the above defined problem is NP-complete we have to show that it is in NP, that another NP-complete problem, say Π, can be reduced to it (i.e., any instance of the problem Π can be transformed into an instance of the *t*-CONSTRAINED ROLE MINING DECISION PROBLEM), and that the reduction can be done in polynomial time. The problem Π used in our simple proof, is the ROLE MINING DECISION PROBLEM (i.e, Problem 2.1).

THEOREM 2.1. *The t-*CONSTRAINED ROLE MINING DECISION PROBLEM *is NP-complete.*

PROOF. The problem is in NP. Indeed the set $\mathcal{R}$ and the matrices URA and RPA constitute a certificate/witness verifiable in polynomial time.
Assume we are given an instance of the ROLE MINING DECISION PROBLEM consisting of a set of users $\mathcal{U}'$, a set of permissions $\mathcal{P}'$, a user-permission assignment $\mathcal{UPA}'$, and a positive integer $k' < \min\{|\mathcal{U}'|, |\mathcal{P}'|\}$. We show how to transform it into an instance of the *t*-CONSTRAINED ROLE MINING DECISION PROBLEM. The reduction is trivial. Indeed, it is enough to set $\mathcal{U} = \mathcal{U}'$, $\mathcal{P} = \mathcal{P}'$, $\mathcal{UPA} = \mathcal{UPA}'$, and $k = k'$ and define

$$t = \max_{u_i \in \mathcal{U}} |\{p_j \in \mathcal{P} : \text{UPA}[i][j] = 1\}|.$$

It is immediate to see that the above reduction can be done in polynomial time and that a solution to the *t*-CONSTRAINED ROLE MINING DECISION PROBLEM directly provides a solution to the ROLE MINING DECISION PROBLEM. Thus, the theorem holds. □

Next we define the optimization version of the *t*-CONSTRAINED ROLE MINING PROBLEM and we show that it is NP-hard.

PROBLEM 2.3. (*t*-CONSTRAINED ROLE MINING OPTIMIZATION PROBLEM) *Given a set of users* $\mathcal{U}$, *a set of permissions* $\mathcal{P}$, *a user-permission assignment* UPA, *and a positive integer* $t$, *what is the smallest integer* $k$ *for which there is a set of roles* $\mathcal{R}$, *a user-to-role assignment* $\mathcal{URA}$, *and a role-to-permission assignment* $\mathcal{RPA}$ *such that* $|\mathcal{R}| = k$, UPA = URA $\otimes$ RPA, *and, for any* $r \in \mathcal{R}$, $|r| \leq t$?

THEOREM 2.2. *The t-*CONSTRAINED ROLE MINING OPTIMIZATION PROBLEM *is NP-hard.*

PROOF. The *t*-CONSTRAINED ROLE MINING OPTIMIZATION PROBLEM is NP-hard, because there exists a trivial polynomial time reduction from the *t*-CONSTRAINED ROLE MINING DECISION PROBLEM to the *t*-CONSTRAINED ROLE MINING OPTIMIZATION PROBLEM. Indeed, we can use an algorithm solving the optimization problem as an oracle to solve the decision problem simply by checking whether the solution of the associated optimization problem has cardinality less than or equal to $k$. □

In [22], Stockmeyer proved that the SET BASIS DECISION PROBLEM is NP-complete by reducing to it the VERTEX COVER DECISION PROBLEM (one of Karp's 21 NP-complete problems [11], see also problem GT1 in [9]). The VERTEX COVER OPTIMIZATION PROBLEM is APX-complete [5], that is, it cannot be approximated within any constant factor in polynomial time unless P=NP. Therefore, we have the following simple non-approximability result:

THEOREM 2.3. *The t-*CONSTRAINED ROLE MINING OPTIMIZATION PROBLEM *cannot be approximated within any constant factor in polynomial time unless P=NP.*

## 3. RELATED WORKS

Role engineering has been firstly introduced by Coyne et al in [4] where the definition of a top down process for the definition of roles has been discussed. Along the same research line, several other works have been presented [18], but recently, the focus of role engineering has turned to consider more automated techniques, based on the bottom up approach, where data mining techniques are applied for the definition of roles [12]. Role mining algorithms have been presented based on set covering [3], graph theory [6, 26], subset enumeration [25], database tiling [23]. The theoretical aspects of the RMP have been considered in [24, 23, 3], where the complexity of the problem has been analyzed and its equivalence to other known optimization problem showed. Another interrelated problem, i.e. dealing with the semantic meaning of roles, has been addressed in [16].

Cardinality constraints on the number of permissions included in a role have been firstly considered in [13], and a heuristic algorithm called Constrained Role Miner (CRM) has been proposed. The CRM algorithm takes in input the UPA matrix and returns a set of roles, each one satisfying the given cardinality constraint. CRM is based on the idea of clustering users having the same set of permissions and selecting, as candidate roles, the roles composed of the set of permissions satisfying the constraint and having the highest number of associated users. In [13], the performances of the algorithm are evaluated on real world datasets considering different metrics (such as the number of returned roles, the sum of the size of the user assignment and permission assignment matrices and so on), with respect to other previously proposed algorithms. However the comparison is performed without considering constraints, since the other algorithms return a complete set of roles but have not the capability of managing constraints. In section 5 we evaluate our proposal against the result obtained after our implementation of the CRM algorithm, considering both real world and synthetic datasets. A different kind of cardinality constraints, considering the number of user-role assignments, have been considered in [10]. Such constraints can be useful when the number of persons that can be assigned to a given role (e.g. the number of directors, managers, etc) in a given organization is known or can be fixed. In the paper, three algorithms have been proposed based on a graph approach, where the role mining problem is mapped to the problem of finding minimum biclique cover of the edges of a bipartite graph. The three heuristics are obtained by modifying the basic greedy algorithm proposed in [6], and experimental results on real world datasets are reported considering some constraints on the number of users that can be assigned to any role. Finally cardinality constraints have also been considered in [15] where a representation of the constraints in terms of association rules is proposed: permissions are regarded as attributes, and each user-permission assignment as a transaction. To generate mutually exclusive permissions constraints, i.e. set of permissions that cannot be assigned to the same role, an algorithm is proposed, based on known techniques for mining association rules in databases, and its performance evaluated on synthetically generated datasets

## 4. HEURISTICS

In this section we present a family of heuristics. Each heuristic takes as input the matrix UPA and returns a complete role set satisfying the cardinality constraint (i.e., at most $t$ permissions are associated to each role). We borrow the ideas from the heuristics presented in [1] and we adapt them to handle the cardinality constraint.

The basic idea is to select from UPA all rows having less than $t$ permissions in an order that will be defined below. Such rows will correspond to candidate roles that will be added to the candidate role-set. If there is no row having at most $t$ permissions, then a row is selected and, $t$ of the permissions included in the row are chosen (the way we select such row and permissions gives rise to different heuristics). The selected permissions induce a role that is added to the candidate role-set. Then, all rows *covered* by the candidate row-set are removed from UPA and the procedure is iterated until the UPA matrix contains some rows.

The above sketched procedure is more formally described by Algorithm 1 where we use the following notation. Given an $a \times b$ binary matrix M, for $1 \leq i \leq a$, with M$[i]$ we denote the M's $i$-th row; while, with $|\texttt{M}[i]|$ we denote the number of ones appearing in M$[i]$. The procedures NUMCOLS(M) and NUMROWS(M), return the number of columns and rows, respectively, of the matrix M. For a set $S$ and an integer $h$, the procedure FIRST($S, h$) returns the first $h$ elements listed in the set $S$. Given a user-permission assignment matrix UPA, a new candidate role is generated by selecting a UPA's row having the least number of permissions with ties broken at random (Lines 6-8 of Algorithm 1). If the number of permissions associated to the selected row (i.e., the number of ones in UPA[$selectedRow$]) is at most $t$, then a new role is created (Line 9 of Algorithm 1). The new role, containing all permissions associated to the selected row, is then added to the candidate role-set (Line 21 of Algorithm 1). In this algorithm, the matrix *uncoveredP* represents the users' permissions that are not covered by the roles in *candidateRoles*. Once we discover a role (i.e, *newRole*) to be added to the *candidateRoles* set, running SETTOZERO (see Algorithm 2) we update the matrix *uncoveredP* according to *newRole*.

All rows whose permissions are covered by the candidate roles are removed from both matrices *uncoveredP* and UPA (Lines 22-23 of Algorithm 1). REMOVECOVEREDUSERS's pseudo-code is quite similar to the pseudo-code for SETTOZERO, hence we omit it[1]. Algorithm 1 halts when all UPA's rows have been removed (Line 5 of Algorithm 1).

If the number of permissions exceeds the cardinality constraint, then two possible ways of selecting the role to be added to the candidate role-set have been considered. These two possibilities gave rise to two heuristics referred to as $t$-SMA$_R$-0 and $t$-SMA$_R$-1, respectively. In $t$-SMA$_R$-0 (i.e., when *selection* is set to 0 in Algorithm 1), the new role will simply contain the first $t$ permissions associated to the selected row (Lines 10-11 of Algorithm 1). While, in $t$-SMA$_R$-1 (i.e., when *selection* is set to 1 in Algorithm 1) we select a row (Lines 13-16 of Algorithm 1) of the matrix *uncoveredP* having the least number of permissions, ties broken at random. In other words, we select a row (i.e, a users) having the least number of permissions still uncovered. If the selected row is associated to more than $t$ permissions, then the new role will only include its first $t$ permissions (Lines 17-19 of Algorithm 1).

---

**Algorithm 2** SETTOZERO(UPA, *uncoveredP*, *newRole*)

1: $np \leftarrow$ NUMCOLS(UPA)
2: $nr \leftarrow$ NUMROWS(UPA)
3: **for** $i = 1$ to $nr$ **do**
4:     $permissions \leftarrow \{p_j : 1 \leq j \leq np \text{ and } \texttt{UPA}[i][j] = 1\}$
5:     **if** $newRole \subseteq permissions$ **then**
6:         **for all** $j$ such that $p_j \in newRole$ **do**
7:             $uncoveredP[i][j] \leftarrow 0$
8:         **end for**
9:     **end if**
10: **end for**
11: Remove from *uncoveredP* all-zeroes rows
12: **return** *uncoveredP*

---

Algorithm 1 returns a set of roles (i.e., rows and subsets of rows) exactly covering the UPA matrix. As described in [1], instead of covering the matrix UPA using its rows we could use its columns. We refer to such a *new* heuristic based on columns as $t$-SMA$_C$. The only difference between heuristics $t$-SMA$_R$ and $t$-SMA$_C$ is the way a role is computed. Heuristic $t$-SMA$_C$ selects a permission $p$ (i.e., a UPA column) associated to the least number of users. Setting $\mathcal{U}(p) = \{u \in \mathcal{U} : (u, p) \in \mathcal{URA}\}$ (i.e, all users having permission $p$), the role $r_p$ induced by permission $p$ is defined as $r_p = \{p' : \mathcal{U}(p) \subseteq \mathcal{U}(p')\} \cup \{p\}$. If $|r_p| \leq t$, then we add it to the candidate role-set; otherwise, we add to the candidate role-set a role comprising the "first" $t$ permissions in $r_p$. As for heuristics $t$-SMA$_R$, rows covered by roles in the candidate role-set are removed. We iterate this process until the UPA matrix contains some rows. $t$-SMA$_C$'s pseudo-code is quite similar to the one for $t$-SMA$_R$, hence we omit it.

## 5. EXPERIMENTAL RESULTS

In this section we evaluate the proposed heuristic by presenting some experimental results obtained executing our

---
[1]Actually, in our implementation SETTOZERO updates both *uncoveredP* and UPA, but in Algorithm 1, for the sake of clarity, we prefer to keep separate the updating of such matrices.

**Algorithm 1** t-SMA$_R$(UPA, $t$, $selection$)
───────────────────────────────────────────
 1: $candidateRoles \leftarrow \emptyset$
 2: $uncoveredP \leftarrow$ UPA
 3: $np \leftarrow$ NUMCOLS(UPA) /* $np$ is equal to the number of permissions */
 4: $nr \leftarrow$ NUMROWS(UPA) /* $nr$ is equal to the number of users */
 5: **while** NUMROWS(UPA) $> 0$ **do**
 6:     $m \leftarrow \min\{|\text{UPA}[i]| : 1 \leq i \leq nr\}$
 7:     $candidateRows \leftarrow \{i : 1 \leq i \leq nr \text{ and } |\text{UPA}[i]| = m\}$
 8:     $selectedRow \leftarrow_R candidateRows$
 9:     $newRole \leftarrow \{p_j : 1 \leq j \leq np \text{ and } \text{UPA}[selectedRow][j] = 1\}$
10:     **if** $|newRole| > t$ and $selection == 0$ **then**
11:       $newRole \leftarrow$ first($newRole, t$)
12:     **else if** $|newRole| > t$ and $selection == 1$ **then**
13:       $m \leftarrow \min\{|uncoveredP[i]| : 1 \leq i \leq nr\}$
14:       $candidateRows \leftarrow \{i : 1 \leq i \leq nr \text{ and } |uncoveredP[i]| = m\}$
15:       $selectedRow \leftarrow_R candidateRows$
16:       $newRole \leftarrow \{p_j : 1 \leq j \leq np \text{ and } uncoveredP[selectedRow][j] = 1\}$
17:       **if** $|newRole| > t$ **then**
18:         $newRole \leftarrow$ first($newRole, t$)
19:       **end if**
20:     **end if**
21:     $candidateRoles \leftarrow candidateRoles \cup \{newRole\}$
22:     $uncoveredP \leftarrow$ setToZero(UPA, $uncoveredP, newRole$)
23:     UPA $\leftarrow$ REMOVECOVEREDUSERS(UPA, $candidateRoles$)
24: **end while**
25: **return** $candidateRoles$
───────────────────────────────────────────

| Dataset | #Users | #Perms | \|UPA\| | min#Perms | max#Perms | Density |
|---|---|---|---|---|---|---|
| Healtcare | 46 | 46 | 1,486 | 7 | 46 | 70.23% |
| Domino | 79 | 231 | 730 | 1 | 209 | 4.00% |
| Emea | 35 | 3,046 | 7,220 | 9 | 554 | 6.77% |
| Firewall1 | 365 | 709 | 31,951 | 1 | 617 | 12.35% |
| Firewall2 | 325 | 590 | 36,428 | 6 | 590 | 19.00% |
| Apj | 2,044 | 1,164 | 6,841 | 1 | 58 | 0.29% |
| Americas large | 3,485 | 10,127 | 185,294 | 1 | 733 | 0.53% |
| Americas small | 3,477 | 1,587 | 105,205 | 1 | 310 | 1.91% |
| Customer | 10,021 | 277 | 45,427 | 0 | 25 | 1.64% |

**Figure 1: Real-world datasets**

heuristics on several input test cases and report some comparisons of their performance to previous proposals. We compare our heuristics with the one described in [13] (from now on denoted CRM). As far as we know, [13] is the only paper to have considered the problem of constructing a role set under cardinality constraints on the roles. In this sense, that is the first comparison between two heuristics dealing with cardinality constraints, since in [13] much of the discussion of the experimental results focused on the comparisons with other heuristics having no limitations on the size of the roles.

The comparison takes into account the metrics introduced in [17]. The goal is to validate our proposal, by showing that its performance, regarding both the execution speed and the *quality* of the returned role set, is equivalent or better than the one returned by CRM. We would like to point out that, using our implementation of CRM, in some cases we obtained different values from the ones presented in [13]. This could be due to different choices in the two implementations (for instance, in our implementation, ties broken at random while it is not clear how they are handled in [13]). Moreover, we had to resolve some ambiguities we found in the description of Algorithm 2 in [13].

All heuristics have been implemented by using Scilab [21] Version 5.3.0 on a MacBook Pro running Mac OS X 10.6.7 (2.66 Ghz Intel Core i7, 4GB 1067Mhz DDR3 SDRAM). In the next section, we compare our heuristics with respect to CRM over available real-world datasets; while, in Section 5.2, we present the results obtained running the implementation of the heuristics over synthetically generated datasets.

## 5.1 Real-world datasets

In this section we compare, on real-world datasets described in Table 1, CRM heuristic with our $t$-SMA$_R$ and $t$-SMA$_C$ heuristics. Such real-world datasets are available online at HP Labs [20] and have been used for evaluation of several other role mining heuristics [6, 17, 13]. The datasets *americas small* and *americas large* have been obtained from Cisco firewalls granting access to the HP network to authenticated users (users' access depends on their profiles). Similar datasets are *apj* and *emea*. The *healthcare* dataset was obtained from the US Veteran's Administration; the *domino* data was from a Lotus Domino server; *customer* is based on the access control graph obtained from the IT department of an HP customer. Finally, the *firewall1* and *firewall2* datasets are obtained as result of executing an analysis algorithm on Checkpoint firewalls. The main characteristics of the nine datasets are reported in Table 1, where we list the number of users and permissions (second and third columns, respectively), the overall number of permissions (i.e., |UPA|), the minimum and maximum number of permissions assigned to a user (sixth and seventh column, respectively), and the UPA's density (i.e., the ratio between |UPA| and the UPA size – #Users × #Perms).

| Dataset | Heuristic | Parameters | | | | | | |
|---|---|---|---|---|---|---|---|---|
| | | $NR$ | \|RH\| | \|URA\| | \|UPA\| | S1 | S2 | CPU time |
| Healtcare | $t$-SMA$_R$ | 16 | 25 | 352 | 429 | 806 | 822 | 0.0107 |
| | $t$-SMA$_C$ | 14 | 23 | 317 | 354 | 694 | 708 | 0.0263 |
| | CRM | 14 | 0 | 317 | 53 | 370 | 384 | 0.0940 |
| Domino | $t$-SMA$_R$ | 20 | 30 | 142 | 627 | 799 | 819 | 0.0176 |
| | $t$-SMA$_C$ | 22 | 42 | 186 | 628 | 856 | 878 | 0.0720 |
| | CRM | 20 | 0 | 177 | 564 | 741 | 761 | 0.1604 |
| Emea | $t$-SMA$_R$ | 34 | 0 | 35 | 7211 | 7246 | 7280 | 0.1425 |
| | $t$-SMA$_C$ | 40 | 20 | 63 | 7514 | 7597 | 7637 | 0.0787 |
| | CRM | 34 | 0 | 35 | 7211 | 7246 | 7280 | 1.8257 |
| Firewall 1 | $t$-SMA$_R$ | 71 | 90 | 2048 | 4398 | 6536 | 6607 | 0.8944 |
| | $t$-SMA$_C$ | 74 | 102 | 3130 | 2800 | 6032 | 6106 | 0.1266 |
| | CRM | 68 | 10 | 2465 | 840 | 3315 | 3383 | 2.6367 |
| Firewall 2 | $t$-SMA$_R$ | 10 | 13 | 836 | 1119 | 1968 | 1978 | 0.0601 |
| | $t$-SMA$_C$ | 10 | 10 | 963 | 998 | 1971 | 1981 | 0.0385 |
| | CRM | 10 | 0 | 963 | 591 | 1554 | 1564 | 0.1246 |
| Apj | $t$-SMA$_R$ | 475 | 304 | 3152 | 2764 | 6220 | 6695 | 39.0043 |
| | $t$-SMA$_C$ | 465 | 320 | 3578 | 2455 | 6353 | 6818 | 4.5203 |
| | CRM | 455 | 3 | 3488 | 1391 | 4882 | 5337 | 184.9613 |
| Americas small | $t$-SMA$_R$ | 225 | 276 | 5045 | 17680 | 23001 | 23226 | 21.7423 |
| | $t$-SMA$_C$ | 204 | 383 | 11936 | 8580 | 20899 | 21103 | 2.5487 |
| | CRM | 209 | 70 | 15580 | 3249 | 18899 | 19108 | 54.4594 |
| Americas large | $t$-SMA$_R$ | 430 | 115 | 3653 | 103541 | 107309 | 107739 | 157.7549 |
| | $t$-SMA$_C$ | 612 | 1647 | 10579 | 84559 | 96785 | 97397 | 24.4408 |
| | CRM | 415 | 32 | 4333 | 87118 | 91483 | 91898 | 796.2600 |
| Customer | $t$-SMA$_R$ | 1154 | 4559 | 46511 | 7519 | 58589 | 59743 | 255.4128 |
| | $t$-SMA$_C$ | 276 | 218 | 45425 | 531 | 46174 | 46450 | 9.3343 |
| | CRM | 277 | 2 | 45443 | 279 | 45724 | 46001 | 425.8314 |

Figure 2: Results of the three heuristics over the real-world datasets

The considered metrics are: the number of roles ($NR$), the size of the role hierarchy (|RH|), the size of the user-to-role assignment matrix (|URA|), the size of the role-to-permission assignment matrix (|RPA|), the sum |URA| + |RPA| + |RH| denoted by $S1$, the size of $NR$ + |URA| + |UPA| + |RH| denoted by $S2$, and the execution time expressed in seconds. This is not at all equivalent to real-world time, but we used those data to compare CPU usage among different heuristics as it is irrespective of background processes that might slow down the execution.

We first tested the heuristics when there is no constraint on the role size (i.e., we set $t$ equal to $max\#Perms$). In this case, in Algorithm 1, setting the parameter *selection* either to 0 or to 1 has no effect on the returned candidate role-set. The results we obtained by running the three heuristics are listed in Figure 2 where both heuristics $t$-SMA$_R$-0 and $t$-SMA$_R$-1 are denoted by $t$-SMA$_R$. Both our heuristics behave pretty well on the nine datasets. Considering the size of the candidate role-set generated by the heuristics, in four cases out of nine (i.e., *Healtcare*, *Domino*, *Emea*, and *Firewall 2*) our heuristics provide the same results as CRM. In four cases out of nine (i.e., *Firewall 1*, *Apj*, *Americas small*, and *Americas large*) CRM returns a (not so much) smaller role-set. Finally, for the *Customer* dataset $t$-SMA$_C$ returns the smallest role-set. Considering the CPU time, our heuristics outperform CRM with improvements ranging from 50% to 90%. If we look at parameters $S1$ and $S2$ we see that, except for the *Emea* dataset, CRM has a better performance. To improve the results we could reduce the size of the role hierarchy RH by running the *lattice-based* postprocessing procedure defined in [6]. According to this procedure, each role $r \in \mathcal{R}$ containing some other roles is substituted with the role $r'$, where

$$r' = r \setminus \bigcup_{r_c\,:\,r_c \subset r} r_c.$$

If $r'$ is empty, then the number of roles is reduced. The substitutions continue until the lattice is completely flat (i.e., no role contains any other role). After running this procedure, the role hierarchy $\mathcal{RH}$ will be empty implying |RH| = 0. The results obtained by running the lattice-based postprocessing processing are shown in Figure 3. Notice that the above mentioned procedure has not been applied when in Figure 2 we have |RH| = 0. As one can see, CRM never computes a smaller role-set than the one returned by our heuristics. Moreover, considering the parameters $S1$ and $S2$, our heuristics in three cases out of nine (i.e., *Healtcare*, *Firewall 2*, and

| Dataset | Heuristic | Parameters | | | | | |
|---|---|---|---|---|---|---|---|
| | | $NR$ | \|URA\| | \|UPA\| | $S1$ | $S2$ | CPU time |
| Healtcare | $t\text{-SMA}_R$ | 14 | 317 | 56 | 373 | 387 | 0.1238 |
| | $t\text{-SMA}_C$ | 14 | 317 | 53 | 370 | 384 | 0.0914 |
| | CRM | 14 | 317 | 53 | 370 | 384 | 0.0940 |
| Domino | $t\text{-SMA}_R$ | 20 | 177 | 564 | 741 | 761 | 0.2637 |
| | $t\text{-SMA}_C$ | 22 | 186 | 358 | 544 | 566 | 0.2101 |
| | CRM | 20 | 177 | 564 | 741 | 761 | 0.1604 |
| Emea | $t\text{-SMA}_R$ | 34 | 35 | 7211 | 7246 | 7280 | 0.1425 |
| | $t\text{-SMA}_C$ | 40 | 63 | 5903 | 5966 | 6006 | 0.7571 |
| | CRM | 34 | 35 | 7211 | 7246 | 7280 | 1.8257 |
| Firewall 1 | $t\text{-SMA}_R$ | 65 | 2552 | 863 | 3415 | 3480 | 4.1333 |
| | $t\text{-SMA}_C$ | 74 | 3130 | 806 | 3936 | 4010 | 1.9595 |
| | CRM | 67 | 2623 | 827 | 3450 | 3517 | 2.6323 |
| Firewall 2 | $t\text{-SMA}_R$ | 10 | 963 | 591 | 1554 | 1564 | 0.1316 |
| | $t\text{-SMA}_C$ | 10 | 963 | 591 | 1554 | 1564 | 0.0853 |
| | CRM | 10 | 963 | 591 | 1554 | 1564 | 0.1246 |
| Apj | $t\text{-SMA}_R$ | 454 | 3489 | 1386 | 4875 | 5329 | 558.9638 |
| | $t\text{-SMA}_C$ | 465 | 3578 | 1347 | 4925 | 5390 | 210.0532 |
| | CRM | 455 | 3497 | 1385 | 4882 | 5337 | 472.8815 |
| Americas small | $t\text{-SMA}_R$ | 197 | 11936 | 2816 | 14752 | 14949 | 137.1257 |
| | $t\text{-SMA}_C$ | 204 | 11936 | 2106 | 14042 | 14246 | 26.7846 |
| | CRM | 202 | 15384 | 2629 | 18013 | 18215 | 114.1122 |
| Americas large | $t\text{-SMA}_R$ | 412 | 4256 | 87268 | 91524 | 91936 | 740.6360 |
| | $t\text{-SMA}_C$ | 612 | 10579 | 23908 | 34487 | 35099 | 512.3025 |
| | CRM | 413 | 4333 | 86794 | 91137 | 91550 | 1062.2914 |
| Customer | $t\text{-SMA}_R$ | 276 | 45425 | 277 | 45702 | 45978 | 5998.957 |
| | $t\text{-SMA}_C$ | 276 | 45425 | 277 | 45702 | 45978 | 48.3644 |
| | CRM | 277 | 45425 | 279 | 45702 | 45978 | 461.7376 |

Figure 3: Results of the of lattice-based postprocessing procedure

*Customer*) achieve the same results as CRM; while, in the remaining six cases they return better results. Sometime, the CPU time needed by our heuristics is larger than the one for CRM but still comparable.

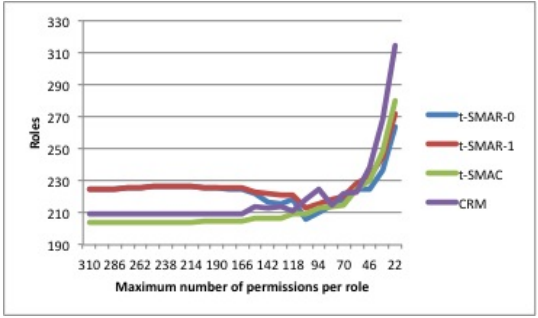

Figure 4: Generated roles for *America small*

We have also evaluated our heuristics by varying the threshold constraint. We performed tests on all the nine datasets, but due to space limitation we only report the results for the *America small* dataset with $t \in \{22 + 12 \cdot i \mid 0 \leq i \leq 24\}$. The reported results (see Figures 4–6) do not include the application of the lattice-based postprocessing procedure. In general, there are few differences between the behavior of $t\text{-SMA}_R$-0 and $t\text{-SMA}_R$-1. Indeed, the graphics associated to them almost overlap. As expected, the number of roles increases (see Figure 4) when the constraint value decreases, i.e. when few permissions can be assigned to each role. Our heuristics always return a smaller role-set than the one computed by CRM. According to Figure 5, the value of the constraint $t$ does not affect much the computation time of our heuristics (the same happens to CRM unless $t < 46$). Anyway, CRM's computation time is 2.5 times larger than the one of $t\text{-SMA}_C$ and about 20 times larger than $t\text{-SMA}_R$-0's ($t\text{-SMA}_R$-1) computation time.

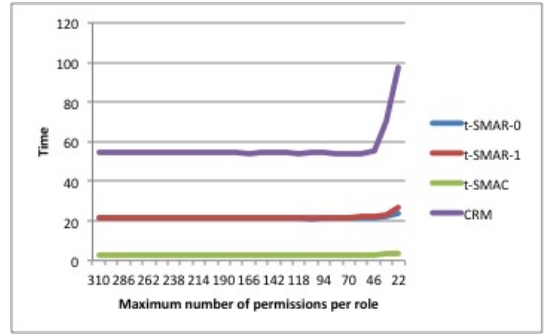

Figure 5: Running time for *America small*

Considering the parameter $S2$, according to Figure , we see that in our heuristics as the value of $t$ decreases, the parameters $S2$ decreases as well. For heuristic CRM, the value of the parameter $S2$ does not change much unless $t < 46$. For $310 \leq t \leq 166$, CRM generate solutions with smaller value of $S2$ with respect to our heuristics; while, for $166 < t \leq 22$, our heuristics have a better performance.

## 5.2 Synthetic datasets

In this section, we report the performance evaluation on synthetic datasets of our heuristics compared to CRM. We followed the approach suggested in [25] generating the datasets by a synthetic data generator. Such a generator takes as input five parameters: the number of roles $NR$, the number of users $NU$, the number of permissions $NP$, the maximal number of roles $MRU$ that can be assigned to each user, and the maximal number of permissions $MPR$ each role can have. To generate the role-to-permission assignment, for each role $r$, the generator chooses a random number $N_r \in [1, MPR]$, then it randomly chooses $N_r$ permissions from $\mathcal{P}$ and assign them to $r$. In this way, we construct the RPA matrix. To obtain the URA matrix, the generator, for each user $u$, chooses a random number $M_r \in [1, MRU]$, then it randomly chooses $M_r$ roles from the generated ones and assign them to $u$. Then, the UPA matrix is implicitly defined.

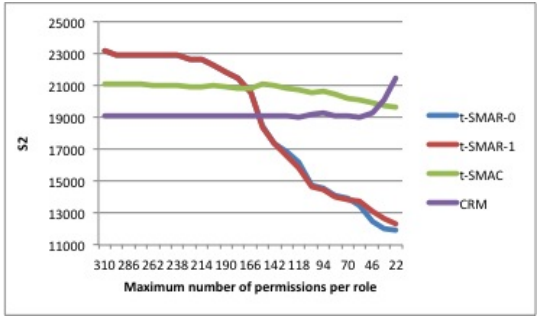

**Figure 6: Value of $S2$ for *America small***

| NR | NU | NP | MRU | MPR |
|---|---|---|---|---|
| 100 | 2000 | 100 | 3 | 10 |
| 100 | 2000 | 500 | 3 | 50 |
| 100 | 2000 | 1000 | 3 | 100 |
| 100 | 2000 | 2000 | 3 | 200 |

**Figure 7: Test parameters with fixed NP/MPR ratio**

We generated datasets using the parameters summarized in Figure 7. As the synthetic data generator is randomized, for each set of parameters, we run it ten times. On each randomly generated dataset (i.e. for each UPA matrix we created) both our heuristics and CRM were run. For a specific parameter set, all reported results are averaged over the ten runs.

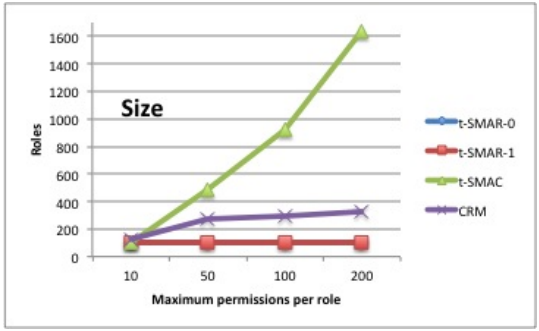

**Figure 8: Role-set size for fixed NP/MPR ratio**

In all our experiments we set the value of the cardinality constraint equal to the maximum number of permissions that can be assigned to each role (i.e., $t = MPR$). We tested the heuristics on several different dataset obtained by keeping constant some parameters while others ranged over different values. For the sake of brevity, we report here only the results of the experiments on the test parameters in Figure 7, where the maximum number of permission per roles ranges from 10 to 200 (and then the same holds for the value of constraint parameter $t$), while the ratio $NP/MRP$ is constant and equal to 10.

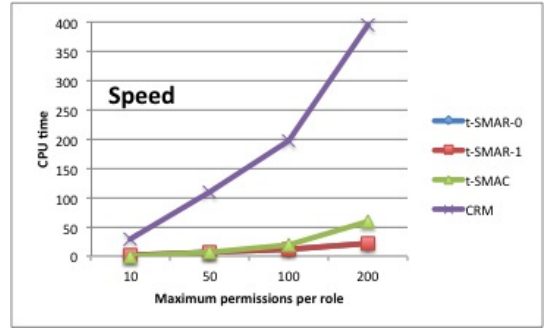

**Figure 9: CPU time for fixed NP/MPR ratio**

For each set of parameters we report the size of the complete role-set generated running the heuristics (see Figure 8) as well as the CPU time need to compute the complete role-set (see Figure 9). We consider also *Accuracy* and *Distance*.

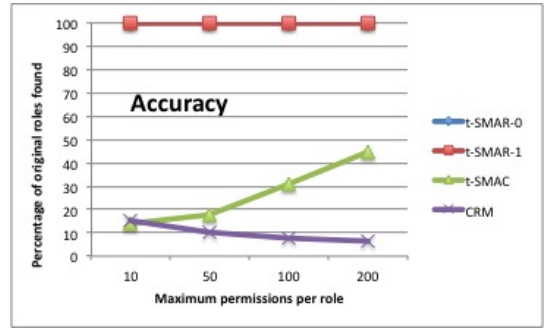

**Figure 10: Accuracy for fixed NP/MPR ratio**

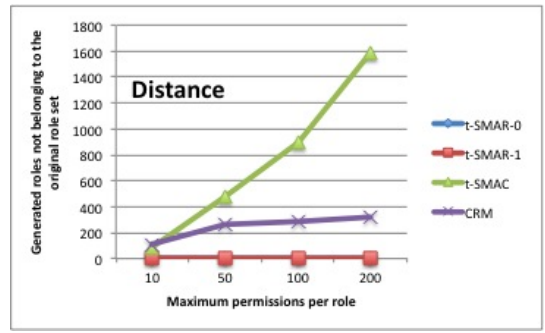

**Figure 11: Distance for fixed NP/MPR ratio**

*Accuracy* (see Figure 10) is defined as the ratio between the number of generated roles exactly matching the *original* roles and the size of role sets generated by the synthetic data generator (i.e., we measure the percentage of original roles found by the heuristics). Given a complete role-set generated by any of the heuristics, the *Distance* parameter (see Figure 11) measures how *different* is the role-set generated by the heuristic from the original one (i.e., if $R_G$ is the role-set generated by the synthetic data generator and $R_F$ is the role-set computed by the heuristic, then $Distance = |R_F \setminus R_G|$). According to data summarized in Figures 8–11, results returned by $t$-SMA$_R$-0 and $t$-SMA$_R$-1 are identical and the graphics associated to them overlap. Both heuristics are much better than CRM. Indeed, they compute in few seconds a role-set having 100% *Accuracy* and null *Distance* meaning that the role-set generate by the synthetic data generator is completely reconstructed for all the considered test cases. Notice that CRM's *Accuracy* is less than 20% and *Distance* and CPU time increases as the maximum number of permissions per role does. Our heuristics $t$-SMA$_R$-0 and $t$-SMA$_R$-1 always return a role-set containing about 100 roles; while CRM returns a much larger role-set (i.e., 4 to 15 times larger).

| MPR | Heuristic | Parameters | | | | | | |
|---|---|---|---|---|---|---|---|---|
| | | $NR$ | \|RH\| | \|URA\| | \|UPA\| | S1 | S2 | CPU time |
| 10 | $t$-SMA$_R$-0 | 99.8 | 59.9 | 5548.6 | 503.4 | 6111.9 | 6211.7 | 1.245 |
| | $t$-SMA$_R$-1 | 99.8 | 59.9 | 5548.6 | 503.4 | 6111.9 | 6211.7 | 1.261 |
| | $t$-SMA$_C$ | 99.6 | 15.3 | 16611.3 | 115.5 | 16742.1 | 16841.7 | 0.299 |
| | CRM | 126.1 | 70.6 | 18274.9 | 171.3 | 18516.8 | 18642.9 | 28.223 |
| 50 | $t$-SMA$_R$-0 | 101.9 | 17.5 | 3673.5 | 2550.4 | 6111.9 | 6343.3 | 5.735 |
| | $t$-SMA$_R$-1 | 102.2 | 21.3 | 3677 | 2550.4 | 6111.9 | 6350.9 | 5.735 |
| | $t$-SMA$_C$ | 490.5 | 502 | 82506.2 | 1035.9 | 84044.1 | 84534.6 | 5.845 |
| | CRM | 269.4 | 383 | 30337.3 | 2292.6 | 18516.8 | 33012.9 | 110.091 |
| 100 | $t$-SMA$_R$-0 | 101.4 | 12.3 | 3622.6 | 4820.7 | 8455.6 | 8557.0 | 11.809 |
| | $t$-SMA$_R$-1 | 101.5 | 13.8 | 3622.9 | 4820.7 | 8457.4 | 8558.9 | 11.807 |
| | $t$-SMA$_C$ | 924.6 | 1674.4 | 152401.2 | 2978.7 | 157054.3 | 157978.9 | 19.383 |
| | CRM | 289.9 | 393.0 | 31625.3 | 4564.0 | 36582.3 | 36872.2 | 195.959 |
| 200 | $t$-SMA$_R$-0 | 100.0 | 3.1 | 3451.1 | 9914.5 | 13368.7 | 13468.7 | 21.838 |
| | $t$-SMA$_R$-1 | 100.0 | 3.1 | 3451.1 | 9914.5 | 13368.7 | 13468.7 | 21.806 |
| | $t$-SMA$_C$ | 1634.0 | 4301.4 | 270813.9 | 7924.0 | 283039.3 | 284673.3 | 59.069 |
| | CRM | 329.5 | 476.1 | 34188.6 | 9741.2 | 44405.9 | 44735.4 | 393.904 |

Figure 12: Results of the four heuristics over synthetic datasets

We also evaluated all heuristics using the metrics considered in Section 5.1. Results are summarized in Figures 12 and 13. We can see that, although in some case CRM returns good value for |UPA|, in general its performances are not very good. We also applied the lattice-based postprocessing procedure to the role-set obtained by running the four heuristics. In Figure 13 we report its results. The procedure improves some of the parameters, flattening, as expected, the role hierarchy and increasing the total CPU time. Even with the postprocessing, the results achieved by CRM are not so accurate. Only for the datasets constructed using the first test parameters (i.e., when $MPR = 10$), CRM returns the original number of roles (with a low *Accuracy*, anyway); while, in the other test cases it remains far from the expected number. Postprocessing has few influence on $t$-SMA$_C$, since the number of roles does not change, while only the |UPA| values improve and in the same measure $S1$ and $S2$, too.

## 6. CONCLUSIONS

The role mining process, usually, returns a role infrastructure on the basis of the relationships among users and permissions contained in the UPA matrix. However, the definition of a role-set really reflecting the internal functionalities of the examined organization remains a challenging task. The need for managing different kind of constraints in role engineering has recently been the focus of many works in literature [13, 10, 15]. The definition and the management of constraints in role mining are very important aspects in role mining, since they allow the role engineer to control the automatic process and introduce some rules that can have impact on the retrieved structure. In this paper, we have proposed a heuristic capable of returning a complete role-set satisfying constraints on the maximum number of permissions included in each role. The comparisons made show how the results in terms of accuracy, distance, size, and computation time improve on a previously presented algorithm [13]. Our simple algorithm is easily extensible to consider other kinds of cardinality constraints, such as maximum number of users assigned to a role or mutually exclusive permissions or roles [15]. Furthermore, it is possible to investigate on the definition of other kinds of constraints regarding the role hierarchy and the semantic associated to each role [16], and try to adapt the proposed algorithm in order to return a role set satisfying the newly defined constraints.

## 7. REFERENCES
[1] C. Blundo and S. Cimato. A simple role mining algorithm. In *Proceedings of the 2010 ACM Symposium on Applied Computing*, SAC '10, pages 1958–1962, New York, NY, USA, 2010. ACM.
[2] A. Bucker. Identity management design guide with ibm tivoli identity manager, 2005.
[3] L. Chen and J. Crampton. Set covering problems in role-based access control. In *ESORICS 2009, Proceedings of 14th European Symposium on Research in Computer Security*, volume 5789 of *LNCS*, pages 689–704. Springer-Verlag, 2009.
[4] E. J. Coyne. Role engineering. In *ACM Workshop on Role-Based Access Control*, 1995.
[5] I. Dinur and S. Safra. On the hardness of approximating minimum vertex cover. *Annals of Mathematics*, 162:2005, 2004.
[6] A. Ene, W. Horne, N. Milosavljevic, P. Rao, R. Schreiber, and R. E. Tarjan. Fast exact and

| MPR | Heuristic | Parameters | | | | | |
|---|---|---|---|---|---|---|---|
| | | $NR$ | $|URA|$ | $|UPA|$ | S1 | S2 | CPU time |
| 10 | $t$-SMA$_R$-0 | 99.7 | 6393.2 | 410.8 | 6804 | 6903.7 | 9.416 |
| | $t$-SMA$_R$-1 | 99.7 | 6393.2 | 410.8 | 6804 | 6903.7 | 9.414 |
| | $t$-SMA$_C$ | 99.6 | 16611.3 | 99.6 | 16710.9 | 16810.5 | 2.201 |
| | CRM | 99.6 | 16611.3 | 99.6 | 16710.9 | 16810.5 | 31.557 |
| 50 | $t$-SMA$_R$-0 | 100.2 | 3691.1 | 2468.8 | 6159.9 | 6260.1 | 10.136 |
| | $t$-SMA$_R$-1 | 100.2 | 3783.9 | 2466.9 | 6250.8 | 6351.3 | 10.155 |
| | $t$-SMA$_C$ | 490.5 | 82506.2 | 502.8 | 83099 | 83499.5 | 212.402 |
| | CRM | 264.3 | 32809.4 | 1733.3 | 34542.7 | 34807 | 222.935 |
| 100 | $t$-SMA$_R$-0 | 100 | 3622.8 | 4720.1 | 8342.9 | 8442.2 | 15.892 |
| | $t$-SMA$_R$-1 | 100.1 | 3628.2 | 4719.9 | 8348.1 | 8448.2 | 15.895 |
| | $t$-SMA$_C$ | 924.6 | 152401.2 | 1109.6 | 153510.8 | 154435.4 | 1393.488 |
| | CRM | 287.2 | 33365.9 | 3946.4 | 37312.3 | 37599.5 | 316.930 |
| 200 | $t$-SMA$_R$-0 | 100 | 3451.1 | 9911.4 | 13362.5 | 13462.5 | 25.284 |
| | $t$-SMA$_R$-1 | 100 | 3451.1 | 9911.4 | 13362.5 | 13462.5 | 25.260 |
| | $t$-SMA$_C$ | 1634 | 270813.9 | 2882.2 | 273696.1 | 275330.1 | 7808.980 |
| | CRM | 327.9 | 35240.2 | 8932.2 | 44172.4 | 44500.3 | 549.864 |

Figure 13: Results of the four heuristics over synthetic datasets after lattice-based postprocessing

heuristic methods for role minimization problems. In *SACMAT '08: Proceedings of the 13th ACM symposium on Access control models and technologies*, pages 1–10. ACM, 2008.

[7] D. F. Ferraiolo, R. S. Sandhu, S. I. Gavrila, D. R. Kuhn, and R. Chandramouli. Proposed NIST standard for role-based access control. *ACM Transaction on Information System Security*, 4(3):224–274, 2001.

[8] M. Frank, D. A. Basin, and J. M. Buhmann. A class of probabilistic models for role engineering. In *ACM Conference on Computer and Communications Security*, pages 299–310. ACM, 2008.

[9] M. R. Garey and D. S. Johnson. *Computers and Intractability, A Guide to the Theory of NP-Completeness*. W.H. Freeman and Company, 1979.

[10] M. Hingankar and S. Sural. Towards role mining with restricted user-role assignment. In *Wireless Communication, Vehicular Technology, Information Theory and Aerospace Electronic Systems Technology (Wireless VITAE), 2011 2nd International Conference on*, pages 1–5, 28 2011-march 3 2011.

[11] R. M. Karp. Reducibility Among Combinatorial Problems. In R. E. Miller and J. W. Thatcher, editors, *Complexity of Computer Computations*, pages 85–103. Plenum Press, 1972.

[12] M. Kuhlmann, D. Shohat, and G. Schimpf. Role mining - revealing business roles for security administration using data mining technology. In *Proceedings of the eighth ACM symposium on Access control models and technologies*, SACMAT '03, pages 179–186, New York, NY, USA, 2003. ACM.

[13] R. Kumar, S. Sural, and A. Gupta. Mining rbac roles under cardinality constraint. In *Proceedings of the 6th international conference on Information systems security*, ICISS'10, pages 171–185, Berlin, Heidelberg, 2010. Springer-Verlag.

[14] N. Li, M. V. Tripunitara, and Z. Bizri. On mutually exclusive roles and separation-of-duty. *ACM Trans. Inf. Syst. Secur.*, 10, May 2007.

[15] X. Ma, R. Li, Z. Lu, and W. Wang. Mining constraints in role-based access control. *Mathematical and Computer Modelling*, 2011.

[16] I. Molloy, H. Chen, T. Li, Q. Wang, N. Li, E. Bertino, S. Calo, and J. Lobo. Mining roles with semantic meanings. In *SACMAT '08: Proceedings of the 13th ACM symposium on Access control models and technologies*, pages 21–30. ACM, 2008.

[17] I. Molloy, N. Li, T. Li, Z. Mao, Q. Wang, and J. Lobo. Evaluating role mining algorithms. In *SACMAT'09*, pages 95–104, 2009.

[18] H. Roeckle, G. Schimpf, and R. Weidinger. Process-oriented approach for role-finding to implement role-based security administration in a large industrial organization. In *Proceedings of the fifth ACM workshop on Role-based access control*, RBAC '00, pages 103–110, New York, NY, USA, 2000. ACM.

[19] R. S. Sandhu, E. J. Coyne, H. L. Feinstein, and C. E. Youman. Role-based access control models. *Computer*, 29:38–47, 1996.

[20] R. Schreiber. Datasets used for role mining experiments. http://www.hpl.hp.com/personal/Robert_Schreiber/.

[21] Scilab Consortium. Scilab: The Free Software for Numerical Computation. http://www.scilab.org/, Ver. 5.3.0 (for Mac OS X 10.6.7).

[22] L. J. Stockmeyer. The minimal set basis problem is NP-complete. Technical Report RC 5431, IBM Research, May 1975.

[23] J. Vaidya, V. Atluri, and Q. Guo. The role mining problem: finding a minimal descriptive set of roles. In *SACMAT '07: Proceedings of the 12th ACM symposium on Access control models and technologies*, pages 175–184. ACM, 2007.

[24] J. Vaidya, V. Atluri, and Q. Guo. The role mining problem: A formal perspective. *ACM Trans. Inf. Syst. Secur.*, 13(3), 2010.

[25] J. Vaidya, V. Atluri, and J. Warner. Roleminer: mining roles using subset enumeration. In *CCS '06*, pages 144–153. ACM, 2006.

[26] D. Zhang, K. Ramamohanarao, and T. Ebringer. Role engineering using graph optimisation. In *SACMAT '07: Proceedings of the 12th ACM symposium on Access control models and technologies*, pages 139–144. ACM, 2007.